\documentclass[prb,showpacs,showkeys]{revtex4}

\usepackage{graphicx}
\usepackage{dcolumn}
\usepackage{bm}
\usepackage{amsmath,amssymb,euscript}

\begin{document}


\title{Composite fermions and quartets in the Fermi-Bose mixture
with attractive interaction between fermions and bosons.}

\author{M.Yu. Kagan}
\email{kagan@kapitza.ras.ru} \affiliation{P.L. Kapitza Institute
for Physical Problems, Kosygin street 2, Moscow, Russia}
\author{I.V. Brodsky}
\affiliation{P.L. Kapitza Institute for Physical Problems, Kosygin
street 2, Moscow, Russia}
\author{D.V. Efremov}
\affiliation{Technische Universit\"{a}t Dresden Institut f\"{u}r
Theoretische Physik, 01062, Dresden}
\author{A.V. Klaptsov}
\affiliation{Russian Research Centre "Kurchatov Institute",
Kurchatov square 1, Moscow, Russia, 123182}

\date{\today}

\begin{abstract}
We consider a model of Fermi-Bose mixture with strong hard-core
repulsion between particles of the same sort and attraction
between particles of different sorts. In this case, besides the
standard anomalous averages of the type $\langle b\rangle$;
$\langle bb\rangle$ and $\langle cc\rangle$, a pairing between
fermion and boson of the type $bc$ is possible. This pairing
corresponds to a creation of composite fermions in the system. At
low temperatures and equal densities of fermions and bosons
composite fermions are further paired in quartets. Our
investigations are important for high-$T_c$ superconductors and in
connection with recent observation of weakly bound dimers in
magnetic traps at ultralow temperatures.
\end{abstract}

\maketitle

\section{Introduction.}

A model of Fermi-Bose mixture is very popular nowadays in
connection with different problems in condensed matter physics
such as high-$T_c$ superconductivity, superfluidity in
${}^3$He-$^4$He mixtures \cite{Bardeen1967}, fermionic
superfluidity in magnetic traps and so on.

In high-$T_c$ superconductivity this model was firstly proposed by
J.Ranninger \cite{Ranninger1985,Chakraverty1998} to describe
simultaneously high transition temperature and short coherence
length of SC pairs on one hand and the presence of well-defined
Fermi-surface on the other. Later on P.W.Anderson
\cite{Anderson1987} reformulated this model introducing bosonic
degrees of freedom (holons) and fermionic degrees of freedom
(spinons), which, according to his ideas, experience in strongly
correlated model a phenomena of spin-charge separation.

Since then a lot of prominent scientists try to prove ideas of
Anderson in the framework of 2D Hubbard and $t-J$ models. In this
context it is necessary to mention first of all the ideas of
Laughlin and Patrick Lee \cite{Laughlin1988,Fetter1989,
Lee_Nagaosa1992,Lee_Nagaosa1998}. These ideas are based on an
anionic picture or on slave boson method. However, even these nice
papers do not contain a rigorous proof of spin-charge separation
in the whole parameter region of the phase diagram of high-$T_c$
superconductors. Moreover, the photoemission experiments
\cite{Ding1996} and numerical calculations of Maekawa and Eder
\cite{Ohta_Maekawa1994} show that at least at low temperatures the
Cooper pairs in high-$T_c$ materials are very much the same as in
ordinary superconductors.

In this paper we show that Fermi-Bose mixture with attractive
interaction between fermions and bosons is unstable towards the
creation of composite fermions $f=bc$. Moreover, for low
temperatures and equal densities of fermions and bosons the
composite fermions are further paired in the quartets $\langle
ff\rangle$. Note that a matrix element $\langle f\rangle=\langle
bc\rangle\neq0$ only for the transitions between the states with
$|N_B;N_F\rangle$ and $\langle N_B-1;N_F-1|$, where $N_B$ and
$N_F$ are numbers of particles of elementary bosons and fermions,
respectively. For superconductive state a matrix element $\langle
ff\rangle\neq0$ only for the transitions between the states with
$|N_B;N_F\rangle$ and $\langle N_B-2;N_F-2|$. Our results are
interesting not only for the physics of high-$T_c$ materials, but
also for Fermi-Bose mixtures in magnetic traps where we can easily
tune the parameters of the system such as the particle density and
the sign and strength of the interparticle interaction
\cite{Hofstetter0204237,Milstein0204334}.

\section{Theoretical model.}

A model of the Fermi-Bose mixture has the following form on a
lattice:
\begin{eqnarray}
      H=&&H_F+H_B+H_{BF},\notag\\
      H_F=&&-t_F\sum\limits_{<i\,j>}c^+_{i\sigma}c_{j\sigma} +
    U_{FF}\sum\limits_in^F_{i\uparrow}n^F_{i\downarrow} -
    \mu_F\sum\limits_{i\,\sigma}n^F_{i\sigma},\notag\\
      H_B=&&-t_B\sum\limits_{<i\,j>}b^+_ib_j +
    \frac{1}{2}\,U_{BB}\sum\limits_in^B_in^B_i -
    \mu_B\sum\limits_{i}n^B_{i},\notag\\
      H_{BF}=&&-U_{BF}\sum\limits_{i\,\sigma}n^B_in^F_{i\sigma}.\label{basic}
\end{eqnarray}

This is a lattice analog of the standard Hamiltonian considered
for example in Ref.~\onlinecite{Efremov_Viverit2002} by Efremov
and Viverit. Here $t_F$ and $t_B$ are fermionic and bosonic
hopping amplitudes, $c^+_{i\sigma}, c_{i\sigma}, b^+_i, b_i$ are
fermionic and bosonic creation-annihilation operators. The Hubbard
interactions $U_{FF}$ and $U_{BB}$ correspond to hard-core
repulsions between particles of the same sort. The interaction
$U_{BF}$ corresponds to the attraction between fermions and
bosons. $W_F=8t_F$ and $W_B=8t_B$ are the bandwidths in 2D.
Finally, $\mu_F$ and $\mu_B$ are fermionic and bosonic chemical
potentials. For the square lattice the spectrums of fermions and
bosons after Fourier transform read: $\xi_{p\sigma}=-2t_F(\cos
p_xd + \cos p_yd) - \mu_F$ for fermions, and $\eta_p=-2t_B(\cos
p_xd + \cos p_yd)- \mu_B$ for bosons, where $d$ is a lattice
constant. In the intermediate coupling case
$W_{BF}/\ln(W_{BF}/T_{0BF})<U_{BF}<W_{BF}$ the energy of the bound
state reads:
\begin{equation}
    |E_b|=\frac{1}{2m_{BF}d^2}
    \frac{1}{\exp\left[\frac{2\pi}{m_{BF}U_{BF}}\right]-1},
\end{equation}
where $m_{BF}=m_Bm_F/(m_B+m_F)$ is an effective mass,
$W_{BF}=4/m_{BF}d^2$ and $T_{0BF}=2\pi n/m_{BF}$. For simplicity
we consider a case of equal densities $n_B=n_F=n$ which is more
relevant for physics of holons and spinons.

Note that in intermediate coupling case a binding energy between
fermion and boson $|E_b|$ is larger than bosonic and fermionic
degeneracy temperatures $T_{0B} = 2\pi n_B/m_B$ and $T_{0F}=2\pi
n_F/m_F\equiv \varepsilon_F$, but smaller than the bandwidths
$W_B$ and $W_F$. In this case a pairing of fermions and bosons
$\langle bc\rangle \neq 0$ takes place earlier (at higher
temperatures) than both Bose-Einstein condensation of bosons (or
bibosons) ($\langle b\rangle \neq 0$ or $\langle bb\rangle \neq
0$) and Cooper pairing of fermions ($\langle cc\rangle \neq 0$).
Note that in the case of a very strong attraction $U_{BF}>W_{BF}$
we have a natural result: $|E_b|=U_{BF}$, and an effective mass
$m^*_{BF}=m_{BF}U_{BF}/W_{BF}\gg m_{BF}$ is additionally enhanced
on the lattice \cite{Nozieres1985}. Note also that the Hubbard
interactions $U_{FF}$ and $U_{BB}$ satisfy the inequalities :
$U_{FF}>W_F/\ln(W_F/|E_{b}|)$ and $U_{BB}>W_B/\ln(W_B/|E_{b}|)$.

\begin{figure}
    \includegraphics{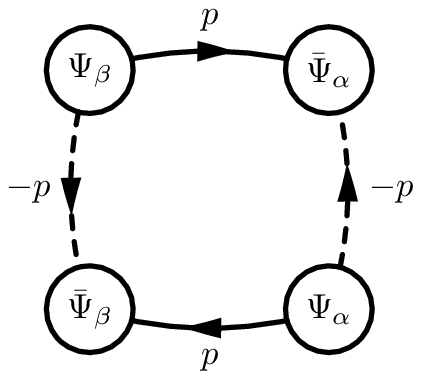}
    \caption{\label{fig:Hubbard-Stratonovich} The skeleton diagram for the coefficient
    $b$ near $\Psi^4$ in the effective action.
    The dashed lines correspond to bosons, the solid lines correspond to fermions.}
\end{figure}

Now let us consider a temperature evolution of the system. It is
governed by the corresponding Bethe-Salpeter equation. After
analytical continuation $i\omega_n\rightarrow \omega+i0$ (see Ref.
\onlinecite{AGD}) the solution of this equation acquires a form:

\begin{equation}\label{Gamma}
    \Gamma(\mathbf{q},\,\omega)= \frac{-U_{BF}}{1 -
    U_{BF}\int\frac{d^2p }{(2\pi)^2}\frac{1-n_F(\xi(\mathbf{p}))
    +n_B(\eta(\mathbf{q-p}))}{\xi(\mathbf{p})+\eta(\mathbf{q-p})-
    \omega-i0}}
\end{equation}
where $\xi(\mathbf{p})=p^2/2m_F-\mu_F;$
$\eta(\mathbf{p})=p^2/2m_B-\mu_B$ are spectrums of fermions and
bosons at low densities $n_Fd^2\ll 1$ and $n_Bd^2\ll 1$. Note that
in the pole of BS-equation enters the temperature factor
$1-n_F(\xi(\mathbf{p}))+n_B(\eta(\mathbf{q-p}))$ in contrast with
the factor $1-n_F(\xi(\mathbf{p}))-n_F(\xi(\mathbf{q-p}))$ for
two-fermion Cooper pairing and $1+n_B(\eta(\mathbf{p}))+
n_B(\eta(\mathbf{q-p}))$ for two-boson pairing. The pole of the
Bethe-Salpeter equation corresponds to the spectrum of the
composite fermions:

\begin{equation}\label{spectr}
    \omega\equiv
    \xi^*_{\mathbf{p}}=\frac{p^2}{2(m_B+m_F)}-\mu_{comp},
\end{equation}
Note that in Eq. (\ref{spectr})
\begin{equation}\label{Chemical_potential}
    \mu_{comp}=\mu_B+\mu_F+|E_b|
\end{equation}
is a chemical potential of composite fermions. Note also that
composite fermions are well-defined quasiparticles, since the
damping of quasiparticles equals to zero in the case of bound
state ($E_b < 0$), but it becomes nonzero and is proportional
$E_b$ in the case of virtual state ($E_b > 0$). The process of a
dynamical equilibrium (boson $+$ fermion $\rightleftarrows$
composite fermion) is governed by the standard Saha formula
\cite{Landau_Stat_One}. In 2D case it reads
\begin{equation}
    \frac{n_Bn_F}{n_{comp}}=\frac{m_{BF}T}{2\pi}\exp\left\{-\frac{|E_b|}{T}\right\}.
\end{equation}
The crossover temperature $T_*$ is defined, as usual, from the
condition that the number of composite fermions equals to the
number of unbound fermions and bosons: $n_{comp}=n_B=n_F=n$. This
conditions yields:
\begin{equation}
    T_*\simeq\frac{|E_b|}{\ln\left(|E_b|/2T_{0BF}\right)}\gg
    \{T_{0B};T_{0F}\}.
\end{equation}

Note that in Boltzmann regime $|E_b|>\{T_{0B};T_{0F}\}$, in fact
we deal with the pairing of two Boltzmann particles. That is why
this pairing does not differ drastically from the pairing of two
particles of the same type of statistics. Indeed, if we substitute
$\mu_B+\mu_F$ in (\ref{Chemical_potential}) on $2\mu_B$ or
$2\mu_F$ we will get the familiar expressions for chemical
potentials of composite bosons consisting either from two bosons
\cite{Nozieres_SaintJames1982,Kagan_Efremov2002} or from two
fermions \cite{Kagan_Beck1998,Kagan_Beck2000}. The crossover
temperature $T_*$ plays the role of a pseudogap temperature, so
the Green functions of elementary fermions and bosons acquire a
two pole structure below $T_*$ in similarity with
Ref.~\onlinecite{Kagan_Beck2000}.

For lower temperatures $T_0<T<T_*$ (where $T_0=2\pi n/(m_F+m_B)$
is degeneracy temperature of composite fermions) the numbers of
elementary fermions and bosons are exponentially small. The
chemical potential of composite fermions reads:
$\mu_{comp}=-T\ln(T/T_0)$. Hence $|\mu_{comp}|\ll |E_b|$ for $T\ll
T_*$.

\begin{figure*}
    \includegraphics{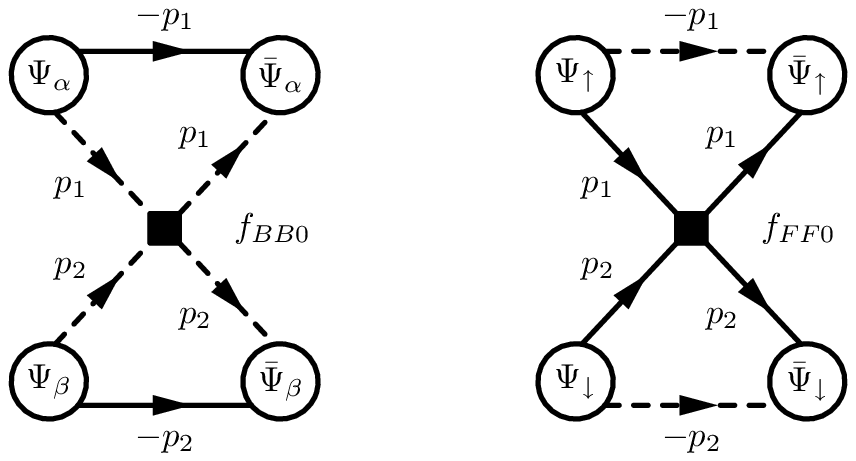}
    \caption{\label{fig:interaction} The corrections to the coefficient
    $b$ containing boson-boson and fermion-fermion interactions.}
\end{figure*}

By performing the Hubbard-Stratonovich  transformation, the
original partition function $Z=\int
\EuScript{D}\bar{b}\,\EuScript{D}b\,\EuScript{D}\bar{c}\,\EuScript{D}
c\exp\left\{-\beta F\right\}$ can be written in terms of the
composite fermions $Z=\int\EuScript{D}\bar{\Psi}_{\alpha}
\EuScript{D}\Psi_{\alpha}\exp\left\{-\beta F_{eff}\right\}$. This
procedure gives the  magnitude of the interaction between the
composite fermions. The lowest order of the series expansion is
given in Fig. \ref{fig:Hubbard-Stratonovich}. Analytically this
diagram is given by:
\begin{equation}\label{integral}
    -\frac{1}{2}\sum\limits_n\int\frac{d^2p}{(2\pi)^2}\,\left\{G_F^2(\mathbf{p};i\omega_{nF})
    G_B^2(-\mathbf{p};-i\omega_{nB})+G_F^2(-\mathbf{p};-i\omega_{nF})
    G_B^2(\mathbf{p};i\omega_{nB})\right\},
\end{equation}
where $G_F=1/(i\omega_{nF}-\xi(\mathbf{p}))$ and
$G_B=1/(i\omega_{nB}-\eta(\mathbf{p}))$ are fermion and boson
Matsubara Green functions, $\omega_{nF}=(2n+1)\pi T$ and
$\omega_{nB}=2n\pi T$ are fermion and boson Matsubara frequencies.
In fact this integral determines the coefficient $b$ near $\Psi^4$
in the effective action. Evaluation of integral (\ref{integral})
yields:
\begin{equation}\label{b}
    b\simeq -N(0)/|E_b|^2,
\end{equation}
where $N(0)=m_{BF}/2\pi$. The corrections to the coefficient $b$
are presented on Fig.~\ref{fig:interaction}. They contain
explicitly the T-matrices for boson-boson and fermion-fermion
interactions. In the intermediate coupling case these diagrams are
small in a small parameters $f_{BB0}\sim1/\ln(W_{B}/|E_{b}|)$ and
$f_{FF0}\sim1/\ln(W_{F}/|E_{b}|)$. So the exchange diagram really
provides the main contribution to the coefficient $b$.

The coefficient near quadratic term $\Psi^2$ in an effective
action in agreement with general rules of diagrammatic technique
(see Ref. \onlinecite{AGD}) is given by:
\begin{equation}\label{a}
    a+cq^2/2(m_B+m_F)=1/\Gamma(q;0),
\end{equation}
where $\Gamma(q;0)$ is given by (\ref{Gamma}). The solution of
(\ref{a}) yields $c=N(0)/|E_b|$, $a=N(0)\ln(T/T_*)$. So in spite
of the fact that in reality $T_*$ corresponds to a smooth
crossover and not to a real second order phase transition, the
effective action of composite fermions at temperatures $T\sim T_*$
formally resembles Ginzburg-Landau functional for Grassman field
$\Psi_{\alpha}$.

If we want to rewrite the effective action with gradient terms
\begin{equation}
    \Delta F=a\bar{\Psi}_{\alpha}\Psi_{\alpha}+
    \frac{c}{2(m_F+m_B)}(\nabla\bar{\Psi}_{\alpha})(\nabla\Psi_{\alpha})
    +\frac{1}{2}b\bar{\Psi}_{\alpha}\bar{\Psi}_{\beta}\Psi_{\beta}\Psi_{\alpha}
\end{equation}
in the form of the energy functional of nonlinear Schr\"{o}dinger
equation for the composite particle with the mass $m_B+m_F$ we
have to introduce the effective order parameter
$\Delta_{\alpha}=\sqrt{c}\,\Psi_{\alpha}$. Accordingly in terms of
$\Delta_{\alpha}$ the new coefficients $\tilde{a}$ and $\tilde{b}$
near quadratic and quartic terms read: $\tilde{a}=a/c$ and
$\tilde{b}=b/c^2$. Note that Grassman field $\Delta_{\alpha}$
corresponds to the composite fermions and is normalized according
to the condition $\Delta^+_{\alpha}\Delta_{\alpha}=n_{comp}$.
Hence the coefficient $\tilde{b}$ plays the role of the effective
interaction between composite particles. From Eqs. (\ref{b}) and
(\ref{a}) $\tilde{b}=-1/N(0)$.

This result coincides by absolute value, but is different in sign
with the results of with Drechsler and Zwerger
\cite{Drechsler1992}, who calculated in 2D case the residual
interaction between two composite bosons each one consisting of
two elementary fermions. The sign difference between these two
results is due to different statistics of elementary particles in
both cases. It is also important to calculate $b(q)$, where the
momenta of the incoming composite fermions equal respectively to
$(\mathbf{q},-\mathbf{q})$. It is easy to find that:
\begin{align}
    b(q)=
    -\frac{1}{2}\sum\limits_n\int\frac{d^2p}{(2\pi)^2}\,\left\{
    G_B(\mathbf{p};i\omega_{nB})G_F(\mathbf{p};-i\omega_{nF})
    G_B(\mathbf{p}+\mathbf{q};i\omega_{nB})
    G_F(\mathbf{p}-\mathbf{q};-i\omega_{nF})\right.+\\
    \left.+G_B(\mathbf{p};-i\omega_{nB})G_F(\mathbf{p};i\omega_{nF})
    G_B(\mathbf{p}-\mathbf{q};-i\omega_{nB})
    G_F(\mathbf{p}+\mathbf{q};i\omega_{nF})\right\},
\end{align}
Straightforward calculation for small $q$ yields in the case of
equal masses $m_B=m_F=m$:
\begin{equation}
    b(q)=-\frac{m}{4\pi(|E_b|+q^2/4m)^2}.
\end{equation}
Accordingly:
\begin{equation}
    \tilde{b}=\frac{b}{c^2}\approx-\frac{4\pi}{m(1+q^2/4m|E_b|)^2},
\end{equation}
where $|E_b|=1/ma^2$. Analogous result in a 3D case was obtained
by Pieri and Strinati \cite{Strinati2000}. Hence, the four
particle interaction has a Yukawa-form in momentum space.
Therefore: $U_4(r)\approx -1/ma^2\sqrt{2r/a}\exp(-2r/a)$
corresponds to an attractive potential with the radius of the
interaction equal to $a/2$. We can calculate now the binding
energy of quartets $|E_4|$. The straightforward calculation
absolutely similar to the calculation of $|E_b|$ yields:
\begin{equation}
   1=\frac{|\tilde{b}|(m_B+m_F)}{2\pi}\int\limits^{2/a}_{0}
   \frac{qdq}{q^2+(m_B+m_F)|E_4|}.
\end{equation}
Hence:
\begin{equation}
   |E_4|=\frac{4}{a^2(m_B+m_F)
   \left[\exp\left(\dfrac{4\pi}{|\tilde{b}|(m_B+m_F)}\right)-1\right]}.
\end{equation}
For equal masses $m_B=m_F$ a coupling constant
$|\tilde{b}|(m_B+m_F)/4\pi=1/2$ and thus:
\begin{equation}
   |E_4|=\frac{2|E_b|}{(e^{1/2}-1)}\approx3|E_b|.
\end{equation}
The process of dynamical equilibrium (composite fermion $+$
composite fermion $\rightleftarrows$ quartet) is again governed by
the Saha formula of the type:
\begin{equation}
    \frac{n^2_{comp}}{n_{4}}=\frac{m_{4}T}{2\pi}\exp\left\{-\frac{|E_4|}{T}\right\}.
\end{equation}
where $m_4=(m_B+m_F)/2$. The number of composite fermions equals
to half a number of quartets $n_4=n_2/2$ for the crossover
temperature:
\begin{equation}
    T^{(4)}_{**}=\frac{|E_4|}{\ln(|E_4|/2T_0)}.
\end{equation}
Below this temperature the quartets of the type $\langle
f_{i\uparrow}b_i;f_{j\downarrow}b_j\rangle$ play the dominant role
in the system. Note that $T^{(4)}_{**}>T_*$, so quartets are
dominant over pairs (composite fermions) in all the temperature
interval. Note also that the quartets are in spin-singlet state.
The creation of spin-triplet quartets is prohibited or at least
strongly reduced by the Pauli principle. The triplet $p$-wave
pairs of composite fermions are possibly created in a strong
coupling case $|E_b|>W$, when the corrections to the coefficient
$b$ given by the diagrams on Fig.~\ref{fig:interaction} are large
and repulsive. However in this case the small parameters are
absent and it is very dificult to control the diagrammatic
expansion.

\section{Three particle problem.}
If we consider a scattering process of an elementary fermion on a
composite fermion, we get a repulsive sign of the interaction
regardless of the relative spin orientation of composite and
elementary fermions. The same result in 3D for scattering of
elementary fermion on dimer consisting of two fermions was
obtained by Shlyapnikov \textit{et al.} \cite{Shlyapnikov0309010}.
However, for a scattering process of elementary boson on a
composite fermion, we get an attractive sign of the interaction.
Moreover, a fourier-component of the three-particle interaction
for $m_B=m_F=m$ reads in 2D case:
\begin{equation}
    U_3(q)=-\frac{8\pi}{m(1+q^2a^2)}
\end{equation}
Hence
\begin{equation}
    U_3(r)\sim-\frac{1}{ma^2}K_0(r/a)\sim-\frac{1}{ma^2}\sqrt{\frac{a}{r}}e^{-r/a}.
\end{equation}
again corresponds to an attractive potential of the Yukawa type,
but now with a range of the interaction equals to $a$. Calculation
of the three-particle bound-state energy yields:
\begin{equation}
   1=\frac{|U_3(0)|}{2\pi}\int\limits^{1/a}_{0}
   \frac{qdq}{q^2/2m_B+q^2/2(m_B+m_F)+|E_3|}.
\end{equation}
Hence for $m_B=m_F=m$:
\begin{equation}
   |E_3|=\frac{3}{4ma^2}
   \frac{1}{\left[\exp\left(\dfrac{3\pi}{m|U_3|}\right)-1\right]}=
   \frac{3|E_b|}{4(e^{\,3/8}-1)}\approx1.65|E_b|.
\end{equation}
The dynamical equilibrium of the type: composite fermion $+$ boson
$\rightleftarrows$ trio is governed by the following Saha formula:
\begin{equation}
    \frac{n_Bn_{comp}}{n_{3}}=\frac{m_{3}T}{2\pi}\exp\left\{-\frac{|E_3|}{T}\right\}.
\end{equation}
where $m_3=m_B(m_B+m_F)/(2m_B+m_F)$. Accordingly, trios dominate
over unbound bosons for temperatures $T<T^{(3)}_{**}$, where:
\begin{equation}
    T^{(3)}_{**}=\frac{|E_3|}{\ln(|E_3|/2T_0)}.
\end{equation}
Note that $T^{(3)}_{**}<T^{(4)}_{**}$, so trios are not so
important as quartets.

As a result for $T<T^{(4)}_{**}$ there are mostly quartets in the
system. The quartets are bose-condensed at the critical
temperature: $T_c=T_0/(8\ln\ln(4/na^2))$ in the case of equal
masses. It is important to note that in Feshbach resonance scheme
\cite{Hofstetter0204237,Milstein0204334,Timmermans1999} we are
usually in the regime $T\sim T_0$, where quartets prev
ail over
trios and pairs. Note also that octets are not formed in the
system because two quartets repel each other due to Pauli
principle in similarity with the results of
Ref.~\onlinecite{Drechsler1992,Haussmann1993}

\section{Conclusions.}

In conclusion we considered an appearance and pairing of composite
fermions in Fermi-Bose mixture with an attractive interaction
between fermions and bosons.

At equal densities of elementary fermions and bosons, the system
is described at low temperatures by a one-component attractive
Fermi-gas for composite fermions and is unstable towards quartets
formation.

The problem which we considered is important for theoretical
understanding of HTSC materials and for the investigation of
Fermi-Bose mixtures of neutral particles at low and ultralow
temperatures. In high-$T_c$ superconductors the role of bosons is
played by holons, the role of fermions is played by spinons.  At
high temperatures spinons and holons are unbound. At lower
temperatures they are bound in composite fermions and, moreover,
the composite fermions are further paired in quartets (singlet
Cooper pairs). The radius of the quartets (the coherence length of
the Cooper pair) is governed by the binding energy of the quartets
$|E_4|$. If $|E_4|$ is larger that $T_0$, then the quartets are
local: $p_Fa<1$. Finally for $T_c=T_0/(8\ln\ln(4/na^2))$ the local
quartets are bose-condensed and the system becomes
superconductive.

Note that we consider a low density limit $|E_b| \gg T_{0}$. In
the opposite case of higher densities $T_{0} \gg |E_b|$,
Bose-Einstein condensation of holons or biholons (see
Ref.~\onlinecite{Lee_Nagaosa1992}, \onlinecite{Kagan_Efremov2002}
and \onlinecite{Lee_Nagaosa1998}) takes place earlier than a
creation of composite fermions and quartets. Such a state can be
distinguished from the ordinary BCS-superconductor by measuring a
temperature dependence of the specific heat and the normal
density.

In Fermi-Bose mixture our investigations enrich superfluid phase
diagram in magnetic traps and are important in connection with
recent experiments, where weakly bound dimers $^6$Li$_2$ and
$^{40}$K$_2$, consisting of two elementary fermions, were observed
\cite{Regal2003,Levi2003}. Note that in a magnetic trap it is
possible to get an attractive scattering length of fermion-boson
interaction with the help of Feshbach resonance
\cite{Timmermans1999}. Note also that even in the absence of the
Feshbach resonance it is experimentally possible now to create
Fermi-Bose mixture with attractive interaction between fermions
and bosons. For example in Ref.~\onlinecite{Roati2002} and
\onlinecite{Modugno2002} such mixture of $^{87}$Rb (bosons) and
$^{40}$K (fermions) was experimentally studied. Moreover, the
authors of Ref.~\onlinecite{Roati2002} and
\onlinecite{Modugno2002} experimentally observed the collapse of
Fermi-gas with a sudden disappearance of fermionic $^{40}$K atoms
when the system enters into the degenerate regime. We cannot
exclude in principle that it is just manifestation of the creation
of the quartets $\langle bc;bc\rangle$ in the system. Note that in
the regime of strong attraction between fermions and bosons the
phase-separation with the creation of larger clusters or droplets
is also possible. Note also that much slower collapse in bose
subsystem of $^{87}$Rb atoms can be possibly explained by the fact
that the number of Rb atoms in the trap is much larger than the
number of K atoms, so after the formation of composite fermions a
lot of residual bosons are still present in the system. The more
thorough comparison of our results with an experimental situation
will be subject of a separate publication. Here we would like to
mention only that for experiments performed in
Ref.~\onlinecite{Roati2002} and \onlinecite{Modugno2002}, a 3D
case is more actual. In the 3D case an attractive interaction
between composite fermions acquires a form
\begin{equation}
       \tilde{b}(q)=-\frac{\pi
       a_{eff}}{m_{BF}[1+q^2/2(m_B+m_F)|E_b|]},
\end{equation}
where $|E_b|=1/2m_{BF}a^2$ is a shallow level of a fermion-boson
bound state. Note that in the case of the repulsive interaction
between two bosons, each one consisting of two fermions,
$a_{eff}=2a$ in the mean-field theory of Haussmann
\cite{Haussmann1993}, $a_{eff}=0.75a$ in the calculations of
Strinati \textit{et al.} \cite{Strinati2000} and $a_{eff}=0.6a$ in
the calculations of Shlyapnikov \textit{et al.}
\cite{Shlyapnikov0309010}. The shallow bound state of quartets
exists in the 3D case only if
\begin{equation}
       a_{eff}>2\pi
       a\left(\frac{m_{BF}}{m_{B}+m_{F}}\right)^{3/2}.
\end{equation}
For $m_{B}=m_{F}=m$: $a_{eff} >\pi a/4$.

\begin{acknowledgments}
    The authors acknowledge helpful discussions with A. Andreev,
    Yu. Kagan,  L. Keldysh, B. Meierovich, P. W\"{o}lfle, G. Khaliullin,
    A. Chernyshev, I. Fomin, M. Mar'enko, A. Smirnov, P. Arseev, E. Maksimov,
    G. Shlyapnikov, M. Baranov and A. Sudb\o. This work was also supported by
    Russian Foundation for Basic Research (Grant No. 02-02-17520), Russian President
    Program for Science Support (Grant No. 00-15-96-9694), and by the
    Grant of Russian Academy of Sciences for Young Scientists.
\end{acknowledgments}

\bibliography{BoseFermi}

\begin{thebibliography}{29}
\expandafter\ifx\csname natexlab\endcsname\relax\def\natexlab#1{#1}\fi
\expandafter\ifx\csname bibnamefont\endcsname\relax
  \def\bibnamefont#1{#1}\fi
\expandafter\ifx\csname bibfnamefont\endcsname\relax
  \def\bibfnamefont#1{#1}\fi
\expandafter\ifx\csname citenamefont\endcsname\relax
  \def\citenamefont#1{#1}\fi
\expandafter\ifx\csname url\endcsname\relax
  \def\url#1{\texttt{#1}}\fi
\expandafter\ifx\csname urlprefix\endcsname\relax\def\urlprefix{URL }\fi
\providecommand{\bibinfo}[2]{#2}
\providecommand{\eprint}[2][]{\url{#2}}

\bibitem[{\citenamefont{Bardeen et~al.}(1967)\citenamefont{Bardeen, Baym, and
  Pines}}]{Bardeen1967}
\bibinfo{author}{\bibfnamefont{J.}~\bibnamefont{Bardeen}},
  \bibinfo{author}{\bibfnamefont{G.}~\bibnamefont{Baym}}, \bibnamefont{and}
  \bibinfo{author}{\bibfnamefont{D.}~\bibnamefont{Pines}},
  \bibinfo{journal}{Phys. Rev.} \textbf{\bibinfo{volume}{156}},
  \bibinfo{pages}{207} (\bibinfo{year}{1967}).

\bibitem[{\citenamefont{Ranninger and Robaszkiewicz}(1985)}]{Ranninger1985}
\bibinfo{author}{\bibfnamefont{J.}~\bibnamefont{Ranninger}} \bibnamefont{and}
  \bibinfo{author}{\bibfnamefont{S.}~\bibnamefont{Robaszkiewicz}},
  \bibinfo{journal}{Physica B} \textbf{\bibinfo{volume}{135}},
  \bibinfo{pages}{468} (\bibinfo{year}{1985}).

\bibitem[{\citenamefont{Chakraverty et~al.}(1998)\citenamefont{Chakraverty,
  Ranninger, and Feinberg}}]{Chakraverty1998}
\bibinfo{author}{\bibfnamefont{B.~K.} \bibnamefont{Chakraverty}},
  \bibinfo{author}{\bibfnamefont{J.}~\bibnamefont{Ranninger}},
  \bibnamefont{and} \bibinfo{author}{\bibfnamefont{D.}~\bibnamefont{Feinberg}},
  \bibinfo{journal}{Phys. Rev. Lett.} \textbf{\bibinfo{volume}{81}},
  \bibinfo{pages}{433} (\bibinfo{year}{1998}).

\bibitem[{\citenamefont{Anderson}(1987)}]{Anderson1987}
\bibinfo{author}{\bibfnamefont{P.~W.} \bibnamefont{Anderson}},
  \bibinfo{journal}{Science} \textbf{\bibinfo{volume}{235}},
  \bibinfo{pages}{1196} (\bibinfo{year}{1987}).

\bibitem[{\citenamefont{Lee and Nagaosa}(1992)}]{Lee_Nagaosa1992}
\bibinfo{author}{\bibfnamefont{P.~A.} \bibnamefont{Lee}} \bibnamefont{and}
  \bibinfo{author}{\bibfnamefont{N.}~\bibnamefont{Nagaosa}},
  \bibinfo{journal}{Phys. Rev. B} \textbf{\bibinfo{volume}{46}},
  \bibinfo{pages}{5621} (\bibinfo{year}{1992}).

\bibitem[{\citenamefont{Lee et~al.}(1998)\citenamefont{Lee, Nagaosa, Ng, and
  Wen}}]{Lee_Nagaosa1998}
\bibinfo{author}{\bibfnamefont{P.~A.} \bibnamefont{Lee}},
  \bibinfo{author}{\bibfnamefont{N.}~\bibnamefont{Nagaosa}},
  \bibinfo{author}{\bibfnamefont{T.~K.} \bibnamefont{Ng}}, \bibnamefont{and}
  \bibinfo{author}{\bibfnamefont{X.~G.} \bibnamefont{Wen}},
  \bibinfo{journal}{Phys. Rev. B} \textbf{\bibinfo{volume}{57}},
  \bibinfo{pages}{6003} (\bibinfo{year}{1998}).

\bibitem[{\citenamefont{Laughlin}(1988)}]{Laughlin1988}
\bibinfo{author}{\bibfnamefont{R.~B.} \bibnamefont{Laughlin}},
  \bibinfo{journal}{Phys. Rev. Lett.} \textbf{\bibinfo{volume}{60}},
  \bibinfo{pages}{2677} (\bibinfo{year}{1988}).

\bibitem[{\citenamefont{Fetter et~al.}(1989)\citenamefont{Fetter, Hanna, and
  Laughlin}}]{Fetter1989}
\bibinfo{author}{\bibfnamefont{A.~L.} \bibnamefont{Fetter}},
  \bibinfo{author}{\bibfnamefont{C.~B.} \bibnamefont{Hanna}}, \bibnamefont{and}
  \bibinfo{author}{\bibfnamefont{R.~B.} \bibnamefont{Laughlin}},
  \bibinfo{journal}{Phys. Rev. B} \textbf{\bibinfo{volume}{39}},
  \bibinfo{pages}{9679} (\bibinfo{year}{1989}).

\bibitem[{\citenamefont{Ding et~al.}(1996)}]{Ding1996}
\bibinfo{author}{\bibfnamefont{H.}~\bibnamefont{Ding}} \bibnamefont{et~al.},
  \bibinfo{journal}{Phys. Rev. B} \textbf{\bibinfo{volume}{54}},
  \bibinfo{pages}{R9678} (\bibinfo{year}{1996}), \bibinfo{note}{and references
  therein}.

\bibitem[{\citenamefont{Ohta et~al.}(1994)\citenamefont{Ohta, Shimozato, Eder,
  and Maekawa}}]{Ohta_Maekawa1994}
\bibinfo{author}{\bibfnamefont{Y.}~\bibnamefont{Ohta}},
  \bibinfo{author}{\bibfnamefont{T.}~\bibnamefont{Shimozato}},
  \bibinfo{author}{\bibfnamefont{R.}~\bibnamefont{Eder}}, \bibnamefont{and}
  \bibinfo{author}{\bibfnamefont{S.}~\bibnamefont{Maekawa}},
  \bibinfo{journal}{Phys. Rev. Lett.} \textbf{\bibinfo{volume}{73}},
  \bibinfo{pages}{324} (\bibinfo{year}{1994}).

\bibitem[{\citenamefont{Hofstetter et~al.}()\citenamefont{Hofstetter, Cirac,
  Zoller, Demler, and Lukin}}]{Hofstetter0204237}
\bibinfo{author}{\bibfnamefont{W.}~\bibnamefont{Hofstetter}},
  \bibinfo{author}{\bibfnamefont{J.~I.} \bibnamefont{Cirac}},
  \bibinfo{author}{\bibfnamefont{P.}~\bibnamefont{Zoller}},
  \bibinfo{author}{\bibfnamefont{E.}~\bibnamefont{Demler}}, \bibnamefont{and}
  \bibinfo{author}{\bibfnamefont{M.~D.} \bibnamefont{Lukin}},
  \eprint{cond-mat/0204237}.

\bibitem[{\citenamefont{Milstein et~al.}()\citenamefont{Milstein, Kokkelmans,
  and Holland}}]{Milstein0204334}
\bibinfo{author}{\bibfnamefont{J.~N.} \bibnamefont{Milstein}},
  \bibinfo{author}{\bibfnamefont{S.~J. J. M.~F.} \bibnamefont{Kokkelmans}},
  \bibnamefont{and} \bibinfo{author}{\bibfnamefont{M.~J.}
  \bibnamefont{Holland}}, \eprint{cond-mat/0204334}.

\bibitem[{\citenamefont{Efremov and Viverit}(2002)}]{Efremov_Viverit2002}
\bibinfo{author}{\bibfnamefont{D.~V.} \bibnamefont{Efremov}} \bibnamefont{and}
  \bibinfo{author}{\bibfnamefont{L.}~\bibnamefont{Viverit}},
  \bibinfo{journal}{Phys. Rev. B} \textbf{\bibinfo{volume}{65}},
  \bibinfo{pages}{134519} (\bibinfo{year}{2002}).

\bibitem[{\citenamefont{Nozieres and Schmittt-Rink}(1985)}]{Nozieres1985}
\bibinfo{author}{\bibfnamefont{P.}~\bibnamefont{Nozieres}} \bibnamefont{and}
  \bibinfo{author}{\bibfnamefont{S.}~\bibnamefont{Schmittt-Rink}},
  \bibinfo{journal}{J. Low Temp. Phys.} \textbf{\bibinfo{volume}{59}},
  \bibinfo{pages}{195} (\bibinfo{year}{1985}).

\bibitem[{\citenamefont{Abrikosov et~al.}(1963)\citenamefont{Abrikosov, Gorkov,
  and Dzyaloshinkski}}]{AGD}
\bibinfo{author}{\bibfnamefont{A.~A.} \bibnamefont{Abrikosov}},
  \bibinfo{author}{\bibfnamefont{L.~P.} \bibnamefont{Gorkov}},
  \bibnamefont{and} \bibinfo{author}{\bibfnamefont{I.~E.}
  \bibnamefont{Dzyaloshinkski}}, \emph{\bibinfo{title}{Methods of Quantum Field
  Theory in Statisitical Physics}} (\bibinfo{publisher}{Dover},
  \bibinfo{address}{New York}, \bibinfo{year}{1963}).

\bibitem[{\citenamefont{Landau and Lifshitz}(1999)}]{Landau_Stat_One}
\bibinfo{author}{\bibfnamefont{L.~D.} \bibnamefont{Landau}} \bibnamefont{and}
  \bibinfo{author}{\bibfnamefont{E.~M.} \bibnamefont{Lifshitz}},
  \emph{\bibinfo{title}{Statistical Physics (Course of Theoretical Physics,
  Volume 5)}} (\bibinfo{publisher}{Butterworth-Heinemann},
  \bibinfo{year}{1999}).

\bibitem[{\citenamefont{Nozieres and {Saint
  James}}(1982)}]{Nozieres_SaintJames1982}
\bibinfo{author}{\bibfnamefont{P.}~\bibnamefont{Nozieres}} \bibnamefont{and}
  \bibinfo{author}{\bibfnamefont{D.}~\bibnamefont{{Saint James}}},
  \bibinfo{journal}{J. Phys. (Paris)} \textbf{\bibinfo{volume}{4}},
  \bibinfo{pages}{1133} (\bibinfo{year}{1982}).

\bibitem[{\citenamefont{Kagan and Efremov}(2002)}]{Kagan_Efremov2002}
\bibinfo{author}{\bibfnamefont{M.~{\ Yu}.} \bibnamefont{Kagan}}
  \bibnamefont{and} \bibinfo{author}{\bibfnamefont{D.~V.}
  \bibnamefont{Efremov}}, \bibinfo{journal}{Phys. Rev. B}
  \textbf{\bibinfo{volume}{65}}, \bibinfo{pages}{195103}
  (\bibinfo{year}{2002}).

\bibitem[{\citenamefont{Kagan et~al.}(1998)\citenamefont{Kagan, Fresard,
  Capezzali, and Beck}}]{Kagan_Beck1998}
\bibinfo{author}{\bibfnamefont{M.~{\ Yu}.} \bibnamefont{Kagan}},
  \bibinfo{author}{\bibfnamefont{R.}~\bibnamefont{Fresard}},
  \bibinfo{author}{\bibfnamefont{M.}~\bibnamefont{Capezzali}},
  \bibnamefont{and} \bibinfo{author}{\bibfnamefont{H.}~\bibnamefont{Beck}},
  \bibinfo{journal}{Phys. Rev. B} \textbf{\bibinfo{volume}{57}},
  \bibinfo{pages}{5995} (\bibinfo{year}{1998}).

\bibitem[{\citenamefont{Kagan et~al.}(2000)\citenamefont{Kagan, Fresard,
  Capezzali, and Beck}}]{Kagan_Beck2000}
\bibinfo{author}{\bibfnamefont{M.~{\ Yu}.} \bibnamefont{Kagan}},
  \bibinfo{author}{\bibfnamefont{R.}~\bibnamefont{Fresard}},
  \bibinfo{author}{\bibfnamefont{M.}~\bibnamefont{Capezzali}},
  \bibnamefont{and} \bibinfo{author}{\bibfnamefont{H.}~\bibnamefont{Beck}},
  \bibinfo{journal}{Physica B} \textbf{\bibinfo{volume}{284-288}},
  \bibinfo{pages}{447} (\bibinfo{year}{2000}).

\bibitem[{\citenamefont{Drechsler and Zwerger}(1992)}]{Drechsler1992}
\bibinfo{author}{\bibfnamefont{M.}~\bibnamefont{Drechsler}} \bibnamefont{and}
  \bibinfo{author}{\bibfnamefont{W.}~\bibnamefont{Zwerger}},
  \bibinfo{journal}{Ann. Phys.} \textbf{\bibinfo{volume}{1}},
  \bibinfo{pages}{15} (\bibinfo{year}{1992}).

\bibitem[{\citenamefont{Pieri and Strinati}(2000)}]{Strinati2000}
\bibinfo{author}{\bibfnamefont{P.}~\bibnamefont{Pieri}} \bibnamefont{and}
  \bibinfo{author}{\bibfnamefont{G.~C.} \bibnamefont{Strinati}},
  \bibinfo{journal}{Phys. Rev. B} \textbf{\bibinfo{volume}{61}},
  \bibinfo{pages}{15370} (\bibinfo{year}{2000}).

\bibitem[{\citenamefont{Petrov et~al.}()\citenamefont{Petrov, Salomon, and
  Shlyapnikov}}]{Shlyapnikov0309010}
\bibinfo{author}{\bibfnamefont{D.~S.} \bibnamefont{Petrov}},
  \bibinfo{author}{\bibfnamefont{C.}~\bibnamefont{Salomon}}, \bibnamefont{and}
  \bibinfo{author}{\bibfnamefont{G.~V.} \bibnamefont{Shlyapnikov}},
  \eprint{cond-mat/0309010}.

\bibitem[{\citenamefont{Timmermans et~al.}(1999)\citenamefont{Timmermans,
  Tommasini, Hussein, and Kerman}}]{Timmermans1999}
\bibinfo{author}{\bibfnamefont{E.}~\bibnamefont{Timmermans}},
  \bibinfo{author}{\bibfnamefont{P.}~\bibnamefont{Tommasini}},
  \bibinfo{author}{\bibfnamefont{M.}~\bibnamefont{Hussein}}, \bibnamefont{and}
  \bibinfo{author}{\bibfnamefont{A.}~\bibnamefont{Kerman}},
  \bibinfo{journal}{Physics Reports} \textbf{\bibinfo{volume}{315}},
  \bibinfo{pages}{199} (\bibinfo{year}{1999}).

\bibitem[{\citenamefont{Haussmann}(1993)}]{Haussmann1993}
\bibinfo{author}{\bibfnamefont{R.}~\bibnamefont{Haussmann}},
  \bibinfo{journal}{Z. Phys. B: Condens. Matter} \textbf{\bibinfo{volume}{91}},
  \bibinfo{pages}{291} (\bibinfo{year}{1993}).

\bibitem[{\citenamefont{Regal et~al.}(2003)\citenamefont{Regal, Ticknor, Bohn,
  and Jin}}]{Regal2003}
\bibinfo{author}{\bibfnamefont{C.~A.} \bibnamefont{Regal}},
  \bibinfo{author}{\bibfnamefont{C.}~\bibnamefont{Ticknor}},
  \bibinfo{author}{\bibfnamefont{J.~L.} \bibnamefont{Bohn}}, \bibnamefont{and}
  \bibinfo{author}{\bibfnamefont{D.~S.} \bibnamefont{Jin}},
  \bibinfo{journal}{Nature} \textbf{\bibinfo{volume}{424}}, \bibinfo{pages}{47}
  (\bibinfo{year}{2003}).

\bibitem[{\citenamefont{Levi}(2003)}]{Levi2003}
\bibinfo{author}{\bibfnamefont{B.~G.} \bibnamefont{Levi}},
  \bibinfo{journal}{Physics Today} \textbf{\bibinfo{volume}{October}},
  \bibinfo{pages}{18} (\bibinfo{year}{2003}).

\bibitem[{\citenamefont{Roati et~al.}(2002)\citenamefont{Roati, Riboli,
  Modugno, and Inguscio}}]{Roati2002}
\bibinfo{author}{\bibfnamefont{G.}~\bibnamefont{Roati}},
  \bibinfo{author}{\bibfnamefont{F.}~\bibnamefont{Riboli}},
  \bibinfo{author}{\bibfnamefont{G.}~\bibnamefont{Modugno}}, \bibnamefont{and}
  \bibinfo{author}{\bibfnamefont{M.}~\bibnamefont{Inguscio}},
  \bibinfo{journal}{Phys. Rev. Lett.} \textbf{\bibinfo{volume}{89}},
  \bibinfo{pages}{150403} (\bibinfo{year}{2002}).

\bibitem[{\citenamefont{Mondugno et~al.}(2002)\citenamefont{Mondugno, Roati,
  Ferlaino, Brecha, and Inguscio}}]{Modugno2002}
\bibinfo{author}{\bibfnamefont{G.}~\bibnamefont{Mondugno}},
  \bibinfo{author}{\bibfnamefont{G.}~\bibnamefont{Roati}},
  \bibinfo{author}{\bibfnamefont{F.}~\bibnamefont{Ferlaino}},
  \bibinfo{author}{\bibfnamefont{R.~J.} \bibnamefont{Brecha}},
  \bibnamefont{and} \bibinfo{author}{\bibfnamefont{M.}~\bibnamefont{Inguscio}},
  \bibinfo{journal}{Science} \textbf{\bibinfo{volume}{297}},
  \bibinfo{pages}{2240} (\bibinfo{year}{2002}).

\end{thebibliography}

\end{document}